\documentclass[conference]{IEEEtran}
\IEEEoverridecommandlockouts

\usepackage{cite}
\usepackage{amsmath,amssymb,amsfonts}
\usepackage{algorithmic}
\usepackage{multirow}
\usepackage{graphicx}
\usepackage{textcomp}
\usepackage{xcolor}
\usepackage{xspace}

\def\BibTeX{{\rm B\kern-.05em{\sc i\kern-.025em b}\kern-.08em
    T\kern-.1667em\lower.7ex\hbox{E}\kern-.125emX}}

\begin{document}
\title{Toward Energy-Efficient and Low-Power Arrhythmia Detection for Wearable Devices}

\author{
    \IEEEauthorblockN{F. Bulten\IEEEauthorrefmark{1}, Y. Rasheed\IEEEauthorrefmark{1}, A. John\IEEEauthorrefmark{2}, V. Stoico\IEEEauthorrefmark{3}, and G.A. Gillani\IEEEauthorrefmark{1}}
    \IEEEauthorblockA{\IEEEauthorrefmark{1}Computer Architecture for Embedded Systems, University of Twente, The Netherlands\\
    \IEEEauthorrefmark{2}Biomedical Signals and Systems, University of Twente, The Netherlands\\
    \IEEEauthorrefmark{3}Faculty of Science, Computer Science, Vrije Universiteit Amsterdam, The Netherlands\\
    Email: f.f.bulten@student.utwente.nl, v.stoico@vu.nl; \{y.rasheed, a.john, s.ghayoor.gillani\}@utwente.nl \thanks{This work is supported by the VU-UT alliance research grant (20007596).}}
}

\IEEEpeerreviewmaketitle
\maketitle
\begin{abstract}
Cardiovascular diseases are the leading cause of death worldwide, and conditions such as arrhythmia often require long-term monitoring for effective detection and diagnosis. However, current wearable monitoring devices are bulky, uncomfortable, and typically rely on clinicians to manually evaluate electrocardiograms (ECGs). While Deep Learning (DL) algorithms have shown superior performance in arrhythmia detection and classification, their computational complexity coupled with high power consumption limit deployment in wearable devices. To address this challenge,
this paper investigates the use of approximation techniques to reduce the power and energy consumption of DL architectures while
maintaining acceptable classification performance. Specifically, techniques such as data precision reduction and approximate multiplication are investigated in a state-of-the-art DL model and its corresponding hardware architecture. The model is trained and validated using the MIT-BIH Arrhythmia Database, and hardware implementations employing various approximate multipliers are synthesized and evaluated. Compared with the state-of-the-art 8.75 $\mu$W (and 2.08 $\mu$J) reference architecture, our proposed architecture consumes 3.07 $\mu$W (and 2.17 $\mu$J) at 12 kHz, showing 64.9\% reduction in power consumption while providing an acceptable output quality, i.e., 93.7\% classification accuracy and 92.1\% sensitivity. At 100 MHz, our proposed architecture consumes 9.45 mW (and 0.8 $\mu$J), showing 61.5\% reduction in energy consumption as compared to the state-of-the-art architecture. These results demonstrate that our proposed approximations significantly extend wearable device battery life while preserving the required arrhythmia classification performance.
\end{abstract}


\section{Introduction}
In 2019, cardiovascular diseases (CVDs) accounted for 32\% of global deaths \cite{CVDwho}. Early detection can significantly improve treatment outcomes; however, some conditions, such as arrhythmias, are unpredictable and require long-term monitoring \cite{Salim}. Electrocardiograms (ECGs) are the standard for monitoring heart activity, but existing wearable and battery-powered devices remain bulky and uncomfortable for prolonged use. For example, Holter monitors \cite{Holtermonitor} record ECG data for 1--2 days before uploading the data to the cloud for clinician review. While effective, this process increases power consumption due to wireless data transfer and requires significant clinician effort for analysis.\\
One approach to reducing clinician workload and communication overhead is to perform automatic CVD classification directly on wearable devices \cite{nguyen2025device}. Deep learning techniques, including deep neural networks (DNNs) and convolutional neural networks (CNNs), have demonstrated strong performance in ECG classification tasks \cite{Ansari}. However, these models are computationally intensive and typically rely on power-hungry hardware, limiting their suitability for battery-powered wearable systems \cite{huynh2017deepmon}.\\
Approximate computing techniques can be used as a means of saving power/energy in neural network inference \cite{armeniakos2022hardware}. However, these methods often degrade classification performance, which is critical in healthcare applications, as any reduction in accuracy can jeopardize patient safety. In domains such as computer vision, these methods have been shown to reduce power consumption effectively without significantly compromising model performance \cite{armeniakos2022hardware}. This is because deep learning models are generally resilient to errors, and these approximations can be accounted for during training, allowing the model to adjust its weights accordingly, resulting in minimal impact on model performance.\\
This paper investigates approximate computing techniques for arrhythmia detection using neural networks based on ECG data. The goal is to maintain a classification sensitivity higher than that of an average cardiologist at 78.4\% \cite{hannun2019cardiologist} while achieving a high efficiency in terms of power/energy. 
The main contributions of this paper are:
\begin{itemize}
   
    \item Investigation of data precision reduction (from 16-bit to 8-bit) and approximate multiplication for a state-of-the-art arrhythmia classification model, which includes approximation-aware retraining.
     \item Investigation of a relatively higher frequency of operation (100 MHz) to improve energy efficiency compared to the state-of-the-art.  
     
    \item Evaluation of the proposed architecture and comparison with the state-of-the-art architectures, based on hardware efficiency and output quality.  

\end{itemize}

\section{Background and Related Work}
\label{literatureReview}
\subsection{ECG classification}
An arrhythmia is an irregularity in the heartbeat. This can occur in a single beat, referred to as morphological arrhythmias, or over multiple heartbeats: rhythmic arrhythmias. This paper focuses on morphological arrhythmias. The Association for the Advancement of Medical Instrumentation (AAMI) standard is used, which suggests categorizing heartbeats into 15 classes with 5 superclasses, normal (N), supra-ventricular ectopic beat (S), ventricular ectopic beat (V), fusion beat (F), and unknown beats (Q) \cite{association1999testing}.
Detecting abnormal heart rhythms manually from ECG recordings is time-consuming for clinicians, particularly because some arrhythmias occur only intermittently during the monitoring period. To address this challenge, automatic ECG classification methods have been developed. A conventional ECG classification pipeline typically consists of four stages: 1) ECG signal processing, 2) heartbeat segmentation, 3) feature extraction, and 4) classification.

Various classification techniques have been proposed in the literature, including Support Vector Machines \cite{SVM}, Linear Discriminants \cite{ResCom}, Reservoir Computing \cite{ResCom}, Dynamically-biased long short-term memory (DBLSTM) \cite{Hu}, and Deep Neural Networks (DNNs) \cite{Janveja}. Among these approaches, DNNs are widely adopted due to their strong classification performance. Unlike traditional machine learning techniques, DNNs can inherently learn relevant features directly from ECG data, eliminating the need for separate handcrafted feature extraction. However, this capability comes at the cost of increased computational complexity and power consumption. Consequently, this paper focuses on enabling deep learning-based ECG classification on low-power wearable devices by combining feature extraction and classification within a single efficient framework.
 
\subsection{Approximation Techniques}
Deep neural networks are naturally resilient to errors, so approximations can be made to improve energy efficiency without significantly compromising overall classification accuracy \cite{Ansari}. Two such methods are precision scaling and approximate arithmetic \cite{armeniakos2022hardware}.

\textit{Precision scaling}: Most DNNs use 32-bit float data precision, however lower precision is a necessity to achieve low power. It has been proven that 8-bit DNN inference can achieve almost identical accuracy performance as its float 32 counterparts, resulting in lower resource usage \cite{Benoit}. One advantage of quantization is that accuracy loss can be tested in software without deployment on hardware. There are two main methods of applying quantization to a DNN: (a) Post-training quantization (PTQ), where the model is quantized after training, and (b) Quantization-aware training (QAT), where quantization is applied at the feed-forward pass during training.

\textit{Approximate Arithmetic}: A widely studied approach is approximate arithmetic \cite{armeniakos2022hardware}, particularly approximate multipliers and adders. Multipliers are preferred due to their higher area and power consumption, offering greater optimization potential. Common types include multiplierless, log-based, compressor-based, recursive, and segmented multipliers. Their performance is evaluated using metrics such as Mean Absolute Error (MAE) and Absolute Mean Error (AME). AME additionally reflects error bias around zero, which is an effective metric to assess MAC operations. 
Multiplierless designs approximate multiplication via shift-and-add operations, reducing power by replacing multipliers with simpler adders.
Log multipliers transform multiplication into addition, saving energy with minimal extra logic \cite{AnsariLog}.
Compressor multipliers generate partial products and combine them using approximate adders instead of exact ones, lowering hardware cost and power \cite{Mannepalli, WarisWang, Kumari}.
Recursive multipliers construct larger multipliers from smaller units \cite{Gillani, Zacharelos, WarisAxrm}, introducing approximation by replacing exact sub-blocks (e.g., \(2 \times 2\)) with approximate versions to reduce complexity. 
Finally, Segmented multipliers save power by omitting computation of some LSBs and using smaller multipliers \cite{Strollo}. Approximate multipliers have been employed for machine-learning-based quality assessment of ECG signals \cite{John_2019}, achieving substantial reductions in power consumption with negligible degradation in accuracy.




The design space for approximate multipliers is large, but certain properties are especially beneficial for neural networks \cite{Ansari}. Since DNNs rely on MAC operations, having both positive and negative errors is advantageous, as they can cancel out during accumulation. Another key approach is approximation-aware training, where approximate multipliers are used during training. This enables the network to adapt its weights accordingly, often improving accuracy and in some cases even outperforming exact multipliers by reducing overfitting \cite{Ansari}.

\section{Methodology}
We investigate precision reduction and approximate multipliers in deep neural networks (DNNs) used for arrhythmia classification to reduce total power consumption without compromising accuracy significantly for application-specific integrated circuits (ASICs).

\textit{Dataset:} A commonly used dataset for arrhythmia classification is the 
MIT-BIH Arrhythmia Database \cite{Moody}.
The MIT-BIH arrhythmia database has 48 half-hour long records sampled at 360 samples/s and 11 bits. The records are filtered with a band pass filter with a bandwidth of 0.1-100 Hz. And all the R-peaks are labeled and associated with one of the 15 standard classes. The dataset is split into the training, validation, and test set in the ratio of 8:1:1.

\begin{figure}
    \centering
    \includegraphics[width=0.9\linewidth]{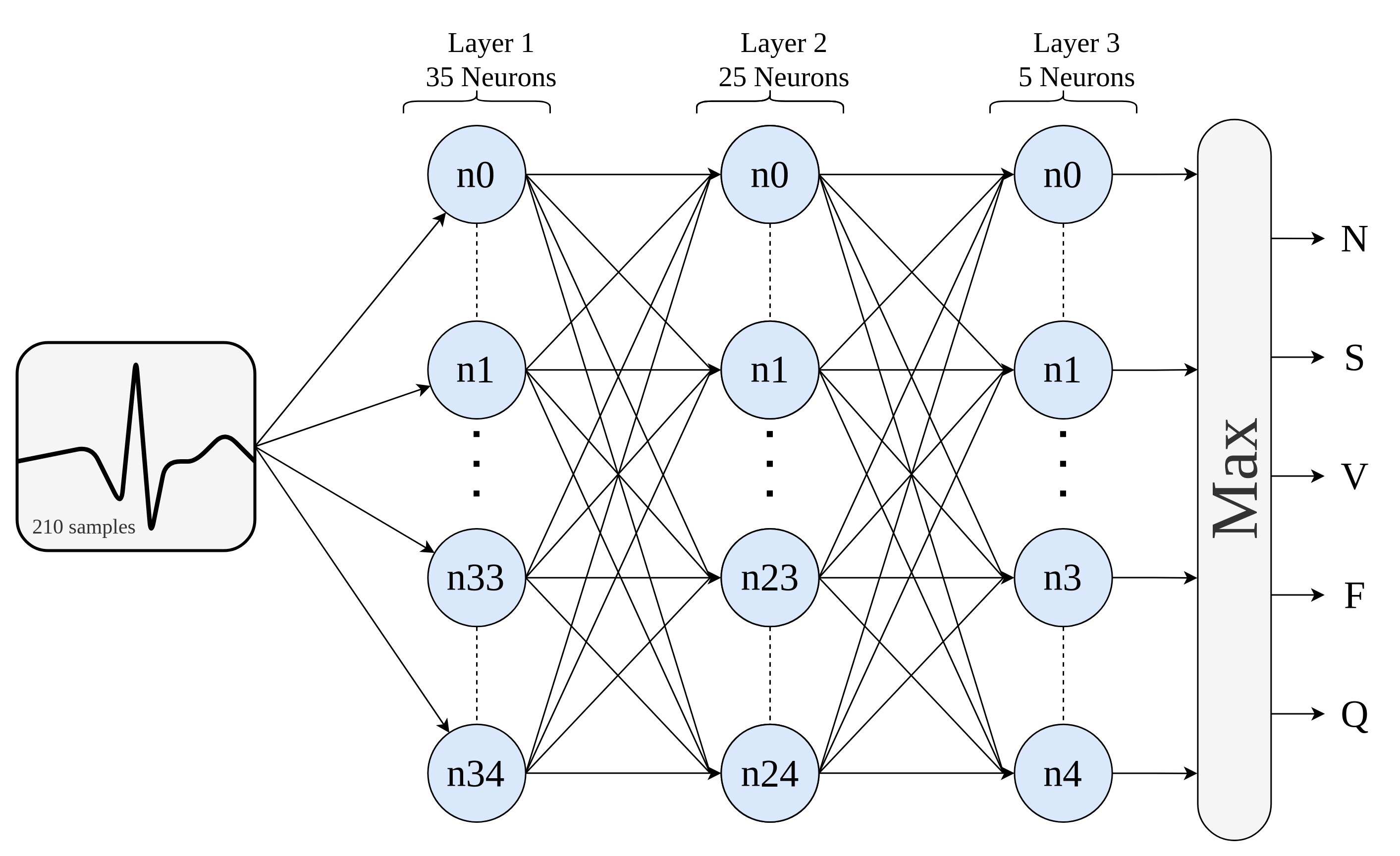}
    \caption{Deep Neural Network (DNN) utilized in the reference design \cite{Janveja}.}
    \label{fig:DNN}
\end{figure}

\textit{Reference Design:}
\label{sec:refdes}
%
The reference model from \cite{Janveja} was chosen because it's a low-power architecture that uses the standard AAMI classes \cite{association1999testing} and could potentially benefit from further approximations, since it currently uses 16‑bit precision and an accurate shift‑add multiplier. The model architecture consists of a peak detection and segmentation module followed by a deep neural network for classification. The model consists of three layers with 35, 25, and 5 neurons, respectively, and takes an input length of 210 samples, as shown in Fig. \ref{fig:DNN}. The neurons in layer 1 and 2 have a ReLU activation function, and layer 3 has a softmax function, returning the index of the largest value. Training is done at float32 accuracy, after which precision is reduced to 16-bit fixed-point with 6 integer bits and 10 decimal bits. Weights and biases are limited from -1 to 1. The model is purposely designed for use in a low-power ASIC. The reference design architecture, as reported in \cite{Janveja}, uses a 16-bit data width and achieves an accuracy of 91.60\%. It operates at a clock frequency of 12 kHz and is implemented in 180 nm SCL technology. The design occupies an area of 1.32 mm\textsuperscript{2} and consumes 8.75 \textmu W of power and 2.08 $\mu$J of energy.


\subsection{Model Training}
Training proceeds in stages from high (16-bit) to low precision weights, as direct low-precision training is ineffective, weight updates smaller than half the quantization step are lost due to post-update quantization. 
To recover performance when using approximate multipliers, the model is first trained with exact multipliers, then retrained using the two most power-efficient 8-bit approximate multipliers from literature as discussed in Section \ref{sec:app_mult}, applied in the forward pass only, with exact multipliers used for backpropagation. Since this retraining is restricted to 8-bit precision, only high learning rates are feasible, producing noisy updates. To mitigate this, training runs for 600 epochs with weights resetting every 15 epochs; the best-performing weights from each window are saved, and the overall best are selected as the final model.

\subsection{Architecture Optimization} 
\begin{figure*}
    \centering
    \includegraphics[width=0.8\linewidth]{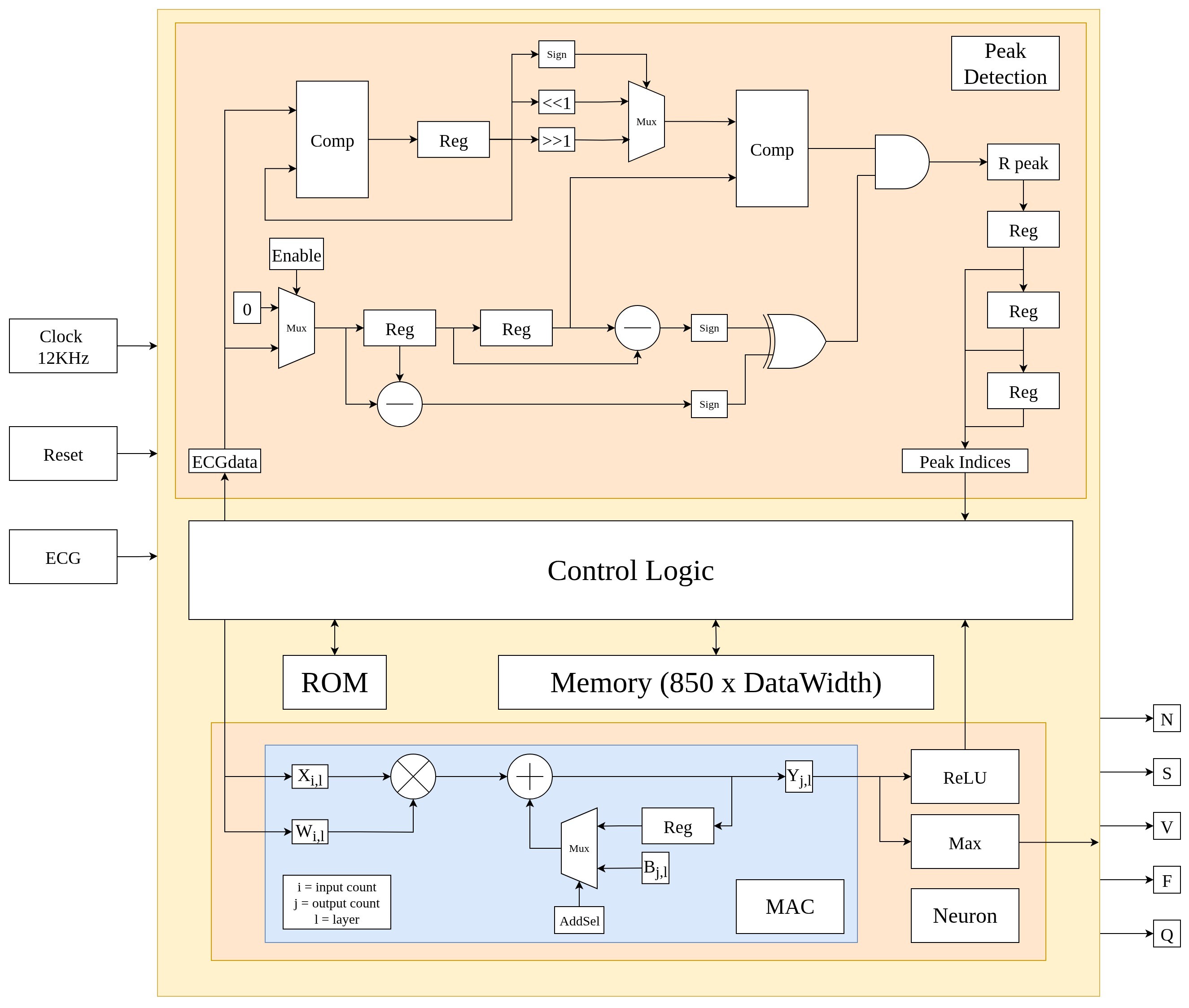}
    \caption{Hardware architecture adapted from \cite{Janveja}. It takes clock, reset, and ECG data as input and provides classification as output.}
    \label{fig:rtl}
\end{figure*}

The hardware architecture can be seen in Fig. \ref{fig:rtl}; the top contains the logic for the peak detection and segmentation, and the bottom has logic for a neuron. The hardware for the neuron is reused for all layers, namely the multiply-accumulate hardware and the hardware for the activation function. Both sections are managed by a controller, which also connects the Read-Only Memory (ROM) and memory to the rest of the logic. The memory is register-based and sized appropriately to store all 850 input samples (each 16 bits wide in the reference architecture and 8 bits wide in ours). ROM is used to store the weights and biases.\\
When the architecture receives data, it is first stored in memory, and the maximum value is determined for peak detection. Once the full ECG window is received, peak detection begins, followed by the classification.
We employed an unsigned multiplier that requires a sign-magnitude representation. In the signed-magnitude format, conversion between positive and negative values only requires flipping the sign bit, which is simpler compared to the two's complement representation. This conversion occurs frequently because all inputs are converted to absolute values before multiplication, and the final result sign is determined using an XOR operation. 

\textit{Peak Detection Hardware:}
As mentioned before, peak detection is done in two phases. During the first phase, it determines the maximum value of the input sample by iterating over the entire input data and storing the result in a register. Afterward, a right or left shift is applied accordingly to compute the threshold. Once iteration is complete and the threshold is found, the peak detection enable signal is set high. The system then iterates over the input data again, this time starting at sample 105 and ending at sample 745. Differences are calculated using registers to obtain delayed samples, and subtractors are used to compute the differences between them. These differences are then compared with the threshold, and their signs are checked. If both conditions are met, the index is stored in a 3-wide shift register. The hardware implementation of this algorithm is shown in the peak detection block in Fig. \ref{fig:rtl}.

\begin{equation}
    \label{eq:neuron}
    y = f_{act} \left(\sum_{i = 0}^{N} X_i w_i + b\right)
\end{equation}

\begin{figure*}
    \centering
    \includegraphics[width=0.8\linewidth]{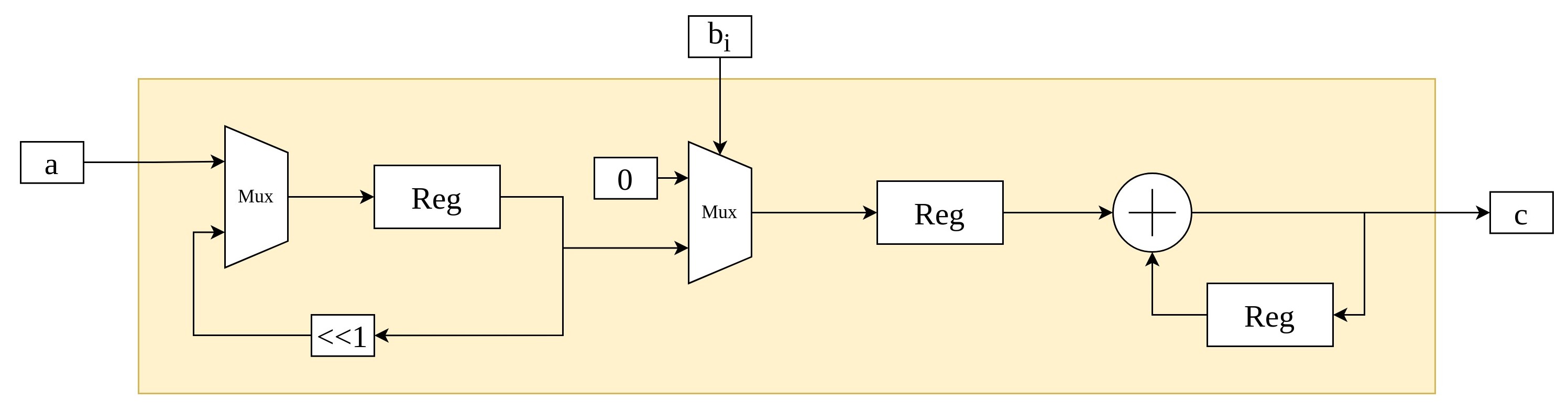}
    \caption{Multiplier design of the reference architecture. Inputs are denoted as a and b, the output as c, and i represents the bit count \cite{Janveja}.}
    \label{fig:refAdd}
\end{figure*}

\textit{Neuron Hardware:}
A neuron computes a weighted sum, as shown in Eq. \ref{eq:neuron}, where x is the input, w is the weight, b is the bias, N is the number of neurons in a layer, and $f_{act}$ is the activation function. This function is implemented in hardware, making use of a MAC unit and two separate activation functions, as shown in the neuron section of Fig. \ref{fig:rtl}. The control logic fetches the appropriate weight or bias, along with the corresponding data. In case of a bias, the input select will be set high, and the bias will be used as input for the adder. Doing it this way will also reset the register, so no additional reset logic is needed. If it is a weight, it will first multiply and add it afterward. In the reference design, the multiplier is a shift-add multiplier with a 32-bit adder, as shown in Fig. \ref{fig:refAdd}. After multiplication and before addition, data is reduced to 16 bits by removing 6 MSBs and 10 LSBs while preserving the sign. After the MAC is done, it goes to the ReLU activation in case of layer 1 or 2, and to the MAX for the last layer. The hardware for the ReLU function becomes a simple MUX selecting the input when the sign is 0, and selecting 0 when the sign is 1. The hardware for the max function contains a register to store the max value and a comparator to check for this value. It takes the output counter to output the predicted class eventually. Fig. \ref{fig:rtl} shows the reference architecture showing ReLU activation; however, we changed this ReLU to the TanH activation. This is done to reduce the chance of overflow in the MAC unit in the subsequent layers. 

\begin{figure}[!htbp]
    \centering
    \includegraphics[width=1\linewidth]{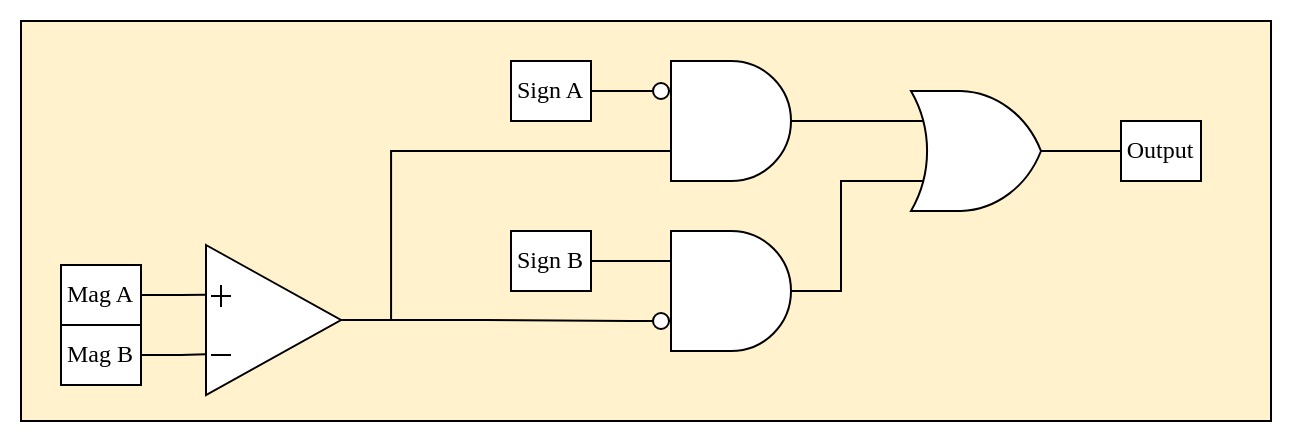}
    \caption{Comparator architecture with inputs A and B.}
    \label{fig:comphard}
\end{figure}

\textit{Sign Magnitude Alterations:}
As mentioned before, sign-magnitude data representation is used with unsigned multipliers. However, this requires suitable comparators, adders, and subtractors. The architecture for the comparator is shown in Fig. \ref{fig:comphard}. In the case of the adder, the truth table is provided in Table \ref{tab:adder table}. When both operands have the same sign, the numbers are added normally. However, when the signs differ, the smaller magnitude is subtracted from the larger, and the output sign corresponds to the sign of the number with the larger magnitude. The subtractors used for peak detection in Fig. \ref{fig:rtl} can remain as standard signed subtractors, since only the sign of the result is used in later logic. However, when the input signs are unequal, the output sign must be explicitly set with the help of the following properties: a positive minus a negative number will always yield a positive result, i.e., $a - b > 0, \text{when } (a > 0) \land (b < 0)$ and a negative minus a positive number will always yield a negative result, i.e.,$a - b < 0, \text{when } (a < 0) \land (b > 0)$. These properties allow the output sign to be directly inferred from the signs of the inputs without the need to compute the full result magnitude beforehand.

\begin{table}[!htbp]
\centering
\caption{Truth Table of the Adder}
\label{tab:adder table}
\begin{tabular}{cccccc}
\hline
\textbf{Sign A} & \textbf{Sign B} & \textbf{mag A} & \textbf{mag B} & \textbf{sign output} & \textbf{mag output} \\ \hline
0 & 0 & - & - & 0 & mag A + mag B \\
0 & 1 & Bigger & - & 0 & mag A - mag B \\
0 & 1 & - & Bigger & 1 & mag B - mag A \\
1 & 0 & Bigger & - & 1 & mag A - mag B \\
1 & 0 & - & Bigger & 0 & mag B - mag A \\
1 & 1 & - & - & 1 & mag A + mag B \\ \hline
\end{tabular}
\end{table}
\subsection{Approximate Multipliers}
\label{sec:app_mult}

Table \ref{tab:my-table} provides an overview of the considered approximate multipliers. The comparison is shown based on power reduction and average mean error (AME) using uniformly distributed 8-bit inputs. From these, six multipliers were selected for investigating approximate multiplication in the model: SSM4, CSSM4, N8-L2, ILM-0, AxRM3, and B81311. Excluded multipliers were either outperformed in power or accuracy (ISH, Ax8) or required multiple clock cycles, increasing latency (PBOM, HOAANED). 

The selected multipliers span a range of trade-offs. SSM4 offers the highest (71.8\%) power reduction (as compared to its accurate counterpart \cite{Strollo})  but has a large error; its corrected variant CSSM4 reduces this bias at modest power cost. N8-L2 provides 45\% power savings with good accuracy. AxRM3 prioritizes accuracy (especially AME) with moderate power savings (21.3\%). ILM-0, a logarithm-based multiplier designed for neural networks, provides 45.9\% power reduction with a near-symmetric error distribution (near-zero AME). The ISH multipliers feature internal self-healing, where positive and negative errors tend to cancel within MAC operations—a property also shared by CSSM4, N8-L2, and AxRM3. 

\begin{table}
\caption{Overview of different 8-bit Approximate Multipliers with state-of-the-art performance}
\label{tab:my-table}
\resizebox{\columnwidth}{!}{%
\begin{tabular}{cccccccc}
\hline
 \textbf{Article} & \textbf{Name}  & \textbf{MAE} & \textbf{AME} & \textbf{Area} \footnotemark{} & \textbf{Power}$^1$ & \textbf{Frequency} & \textbf{Process} \\ \hline
 \cite{Gillani} & Accurate & 0 & 0 & 519 & 439 & 1 GHz & 40nm TSMC \\ 
 & ISH1 & 1438 & 1437 & 415 & 356 & 1 GHz & 40nm TSMC \\ 
 & ISH2 & 1342 & 1334 & 464 & 373 & 1 GHz & 40nm TSMC \\ \hline
 \cite{Mannepalli}& Accurate & 0 & 0 & 404 & 57.43 & 250 MHz & 45nm CMOS \\ 
 & HOAANED & 11.52 & 4.24 & 315 & 47.58 & 250 MHz & 45nm CMOS \\ \hline
\  \cite{AnsariLog}& Accurate & 0 & 0 & 235.9 & 99.3 & 250 MHz & 28nm CMOS \\ 
 & ILM-0 & 458 & 0.47 & 255 & 53.72 & 250 MHz & 28nm CMOS \\ \hline
\cite{Zacharelos}& Accurate & 0 & 0 & 81.45 & 109.38 & 1 GHz & 14nm FinFet \\ 
 & N8-L1 & 495 & 376 & 45.86 & 60.84 & 1 GHz & 14nm FinFet \\ 
 & N8-L2 & 173 & 139 & 43.79 & 59.11 & 1 GHz & 14nm FinFet \\ \hline
\cite{WarisAxrm}& Accurate & 0 & 0 & 185 & 160 &  & 28nm CMOS \\ 
 & AxRM1 & 16 & 0.06 & 171 & 150 &  & 28nm CMOS \\ 
 & AxRM2 & 266 & 0.7 & 159 & 136 &  & 28nm CMOS \\ 
 & AxRM3 & 320 & 50 & 144 & 126 &  & 28nm CMOS \\ \hline
\cite{WarisWang}& Accurate & 0 & 0 & - & 186* &  & 45nm \\ 
 & Ax8-1 & 16 & 3 & 419 & 182 &  & 45nm \\ 
 & Ax8-2 & 397 & 390 & 323 & 157 &  & 45nm \\ 
 & Ax8-3 & 781 & 779 & 302 & 144 &  & 45nm \\ \hline
 \cite{Kumari}& Accurate & 0 & 0 & 482 & 52.57 &  & 40nm UMC \\ 
 & PBOM30 & 0.82 & 0.82 & 437 & 48 &  & 40nm UMC \\ 
 & PBOM50 & 8.43 & 8.43 & 395 & 43 &  & 40nm UMC \\ 
 & PBOM33 & 17 & 17 & 381 & 42 &  & 40nm UMC \\ 
 & PBOM53 & 24 & 24 & 331 & 36 &  & 40nm UMC \\ 
 & PBOM73 & 56 & 56 & 274 & 30 &  & 40nm UMC \\ 
 & PBOM75 & 166 & 166 & 241 & 26 &  & 40nm UMC \\ \hline
 \cite{Strollo}& Accurate & 0 & 0 & 196 & 177 &  & 28nm CMOS \\ 
 & SSM4 & 1743 & 1743 & 66.5 & 49.9 & 1 GHz & 28nm CMOS \\ 
 & CSSM4 & 586 & 279 & 71.2 & 54.52 & 1 GHz & 28nm CMOS \\ 
 & SSM5 & 784 & 784 & 101.5 & 82.48 & 1 GHz & 28nm CMOS \\ \hline
\end{tabular}
}
\footnotemark[\value{footnote}] 
Area is presented in $\mu$m$^2$ and Power is presented in  $\mu$W.\\ * Power has been estimated based on the data provided in \cite{WarisWang}.
\end{table}

\section{Results}
\label{sec:results}
\subsection{Software Results}
The model is trained on 8-bit precision. A significant proportion of the weights are observed to be either at their maximum or minimum representable values—specifically, 0.9921875 and -0.9921875 for 8-bit representation. These correspond to the positive and negative numbers in hardware, which is the limit for 8-bit sign-magnitude representation. This also shows that it might be beneficial to pick approximate multipliers that can accurately multiply by these numbers. This also occurs in the 8-bit layer inputs; because of the TanH function, the layer 1 input is different since this is the ECG data. 

\subsubsection{Model Accuracy}

The confidence matrix for the model when the labeled peaks are used for arrhythmia classification is shown in Table \ref{tab:conf8bit}, where operations are carried out in 8-bit. Here, `N' stands for the normal sinus rhythm, 'S' for supraventricular ectopic beats, 'V' for ventricular ectopic beats, 'F' for fusion beats and 'Q' for unclassified beats. The sensitivities are provided per class and the model accuracy is depicted in the table. 
For the labeled peaks, the model is validated using the full test set, with ECG samples segmented based on the manually annotated peaks in the MIT-BIH dataset. The classification sensitivity surpasses 78.4\%, the reported average sensitivity of a cardiologist \cite{hannun2019cardiologist}. 





\begin{table}
\centering
\caption{confidence matrix for the 8-bit model}
\label{tab:conf8bit}
\begin{tabular}{c|ccccc|c}
\hline
           & \textbf{N} & \textbf{S} & \textbf{V} & \textbf{F} & \textbf{Q} & \textbf{Sensitivity (\%)} \\ \hline
\textbf{N} & 8159 & 195 & 80 & 73 & 37 & 95.49 \\
\textbf{S} & 26 & 177 & 2 & 1 & 1 & 85.51 \\
\textbf{V} & 9 & 8 & 395 & 9 & 2 & 93.39\\
\textbf{F} & 1 & 0 & 4 & 59 & 0 & 92.19\\
\textbf{Q} & 2 & 2 & 3 & 0 & 754 & 99.08\\ \hline \hline
\multicolumn{6}{c|}{\textbf{Accuracy (\%)}}        & 95.44 \\ \hline
\end{tabular}

\end{table}

\subsubsection{Peak Detection Accuracy}
The accuracy of the peak detection algorithm can be seen in Table \ref{tab:peakDetAcc}, in which no peaks were detected in 49 samples and the rest were incorrect peaks.

\begin{table}
\centering
\caption{peak detection accuracy}
\label{tab:peakDetAcc}
\begin{tabular}{ccccc}
\hline
\textbf{Model} & \textbf{Total Peaks} & \textbf{Detected Peaks} & \textbf{Correct Peaks} & \textbf{Percentage} \\ \hline
8-bit & 8000 & 7951 & 7202 & 90.03\% \\ \hline
\end{tabular}
\end{table}

\subsubsection{Model Accuracy with Peak Detection:}
The model was subsequently evaluated using peak-based segmentation,  and the corresponding confusion matrix is provided in Table \ref{tab:conf8bitpd}. The sensitivity remains above the average cardiologist level of 78.4\% \cite{hannun2019cardiologist}. Notably, when comparing the confusion matrices in \ref{tab:conf8bit} and \ref{tab:conf8bitpd}, the most significant improvement is seen in the S class sensitivity, which increases from 85.51\% to 92.68\%.


\begin{table}
\centering
\caption{confidence matrix for the 8-bit model with peak detection}
\label{tab:conf8bitpd}

\begin{tabular}{c|ccccc|c}
\hline
           & \textbf{N} & \textbf{S} & \textbf{V} & \textbf{F} & \textbf{Q} & \textbf{Sensitivity (\%)} \\ \hline
\textbf{N} & 6863       & 121        & 106        & 82         & 11         & 95.54                     \\
\textbf{S} & 8          & 114        & 0          & 1          & 0          & 92.68                     \\
\textbf{V} & 1          & 1          & 22         & 0          & 0          & 91.67                     \\
\textbf{F} & 0          & 0          & 0          & 0          & 0          & 0.00                         \\
\textbf{Q} & 2          & 3          & 1          & 0          & 615        & 99.03 \\ \hline  
\hline
\multicolumn{6}{c|}{\textbf{Accuracy (\%)}}        & 95.76 \\ \hline             
\end{tabular}

\end{table}

\subsection{Hardware Results}
\begin{figure}
\centering
\includegraphics[width = 1\linewidth]{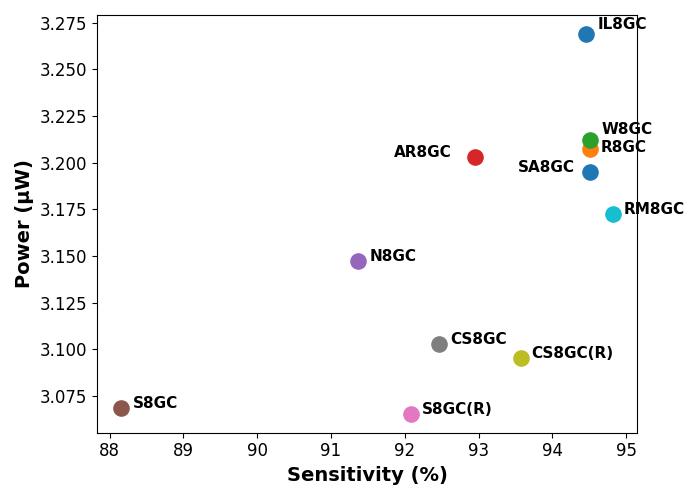}
\caption{Power-sensitivity trade-off for the considered 8-bit architectures. S8GC(R) provides the highest power efficiency with an acceptable output quality.} 
\label{fig:PSres}
\end{figure}

\begin{figure}
\centering
\includegraphics[width = 1\linewidth]{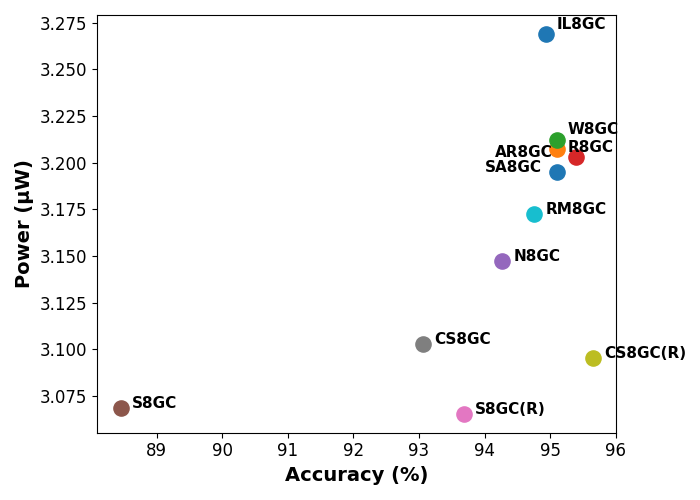}
\caption{Power-accuracy trade-off for the considered 8-bit architectures. Again, S8GC(R) provides the highest power efficiency with an acceptable output quality.} 
\label{fig:PAcc}
\end{figure}

A comparison of sensitivity and accuracy versus power usage across all architectures is shown in Fig. \ref{fig:PSres} and Fig. \ref{fig:PAcc} respectively, with the complete results provided in Table \ref{tab:hard8acc}. Both sensitivity and accuracy are used as evaluation metrics because the dataset is imbalanced, with the majority of signals belonging to the normal class. Consequently, accuracy alone is not representative of classification performance.
 Of the accurate architectures, SA8GC (using shift add multiplier) is the most power-efficient, followed by R8GC (using recursive multiplier) and W8GC (using Wallace multiplier). Without retraining, the most efficient of the approximate architectures is S8GC (using SSM4  multiplier), but this architecture also gives the biggest loss in classification performance. CS8GC architecture (using CSSM4 multiplier) follows in power usage, and has better sensitivity than N8GC (using N8-L2 multiplier), which consumes more power. Moreover, RM8GC architecture (using AxRM3 multiplier) and the IL8GC architecture (using ILM-0 multiplier) have better output quality as compared to S8GC, but they consume more power. Lastly, AR8GC (using B8M1311 multiplier) consumes less power than its accurate counterpart (R8GC), but consumes more power than SA8GC (using the accurate shift add multiplier).

\textit{Retrained Models:}
Table \ref{tab:hard8acc} shows that the S8GC and CS8GC architectures use the least amount of power, so these have been chosen for retraining. The results of the retrained models are also shown in Table \ref{tab:hard8acc}. Compared to the S8GC architecture, the accuracy of the S8GC(R) increases to 93.68\%, and the sensitivity increases to 92.08\%. A minor difference in power usage is noticeable, which is a result of change in weights (different weights cause different switching activity). Similarly, compared to the CS8GC architecture, the accuracy of the CS8GC(R) increases to 95.65\% and the sensitivity increases to 93.57\%.

\textit{Our Proposed Architecture:}
 S8GC(R), CS8GC(R), and RM8GC architectures dominate in the power-sensitivity trade-off, see Fig \ref{fig:PSres}. Similarly for power-accuracy trade-off,  S8GC(R) and CS8GC(R) dominate, see Fig. \ref{fig:PAcc}. In both cases, S8GC(R) provides the highest power efficiency with an acceptable output quality. Therefore, we propose the S8GC(R) architecture for low-power arrhythmia classification. 

\begin{table*}
\caption{8-bit Model Software and Hardware results. SA8GC, R8GC, and W8GC utilize accurate multipliers. S8GC(R) and CS8GC(R) are the retrained architectures. Each architecture has a latency of 8480 clock cycles.}
\label{tab:hard8acc}
\centering
\begin{tabular}{lccccccccccc}
\hline
\textbf{Name} & \textbf{SA8GC} & \textbf{R8GC} & \textbf{W8GC} & \textbf{AR8GC} & \textbf{N8GC} & \textbf{S8GC} & \textbf{S8GC(R)} & \textbf{CS8GC} & \textbf{CS8GC(R)} & \textbf{RM8GC} & \textbf{IL8GC} \\ \hline
\textbf{Multiplier} & Shift Add & Recursive & Wallace & B8M1311 & N8L2 & SSM4 & \textbf{SSM4} & CSSM4 & CSSM4 & AxRM3 & ILM0 \\ 
\textbf{Accuracy (\%)} & 95.10 & 95.10 & 95.10\% & 95.39 & 94.27 & 88.45 & \textbf{93.68} & 93.06 & 95.65 & 94.76 & 94.93 \\ 
\textbf{Sensitivity (\%)} & 94.51 & 94.51 & 94.51 & 92.95 & 91.37\ & 88.16 & \textbf{92.08} & 92.47 & 93.57 & 94.82 & 94.46 \\ 
\textbf{Frequency (kHz)} & 12 & 12 & 12 & 12 & 12 & 12 & \textbf{12} & 12 & 12 & 12 & 12 \\ 
\textbf{Power ($\mu$ W)} & 3.1947 & 3.2072 & 3.2123 & 3.2029 & 3.1472 & 3.0686 & \textbf{3.0654} & 3.1029 & 3.0952 & 3.1726 & 3.2687 \\ 
\textbf{Energy/Class ($\mu$J)} & 2.2576 & 2.2664 & 2.2700 & 2.2634 & 2.2240 & 2.1685 & \textbf{2.1656} & 2.1927 & 2.1873 & 2.2421 & 2.3099 \\ 
\textbf{Area ($mm^2$)} & 1.0358 & 1.0361 & 1.0362 & 1.0360 & 1.0335 & 1.0313 & \textbf{1.0342} & 1.0332 & 1.0358 & 1.0350 & 1.0369 \\ 
\textbf{Slack ($\mu$s)} & 41.665 & 41.665 & 41.665 & 41.665 & 41.665 & 41.665 & \textbf{41.665} & 41.665 & 41.665 & 41.665 & 41.665 \\ \hline
\end{tabular}
\end{table*}

\textit{High Frequency:}
There is a high slack left over, around 41.665 $\mu$s in all cases, meaning frequency could be further increased. For our proposed S8GC(R) architecture, we investigated increase in frequency until the slack becomes 0 ns. It was achieved at 100 MHz. At 100 MHz, our proposed S8GC(R) architecture consumes 9.448 mW power, 0.801 \textmu J energy/classification, 1.028 mm$^2$ chip area, and 8480 clock cycles latency. The improvement is in the energy per classification, dropping from 2.08\textmu J (12 kHz) to 0.80\textmu J (100 MHz). 


\subsection{Comparison with State-of-the-art}
In Table \ref{tab:litcomp}, our proposed S8GC(R) architecture is compared to the state-of-the-art architectures from literature. The S8GC(R) architecture has been chosen since it provides the best trade-off between power savings and output quality, i.e., being the most power-efficient architecture with an acceptable  output quality in terms of classification accuracy and sensitivity. It may be noted that our proposed architecture outperforms all other architectures when it comes to power usage. Compared to the reference architecture \cite{Janveja}, the proposed S8GC(R) architecture provides 64.9\% reduction in power consumption, with better classification accuracy and sensitivity.

\begin{table*}[]
\centering
\caption{Efficiency and performance comparison of the proposed S8GC(R) architecture with that of the state-of-the-art architectures.}
\label{tab:litcomp}
\begin{tabular}{|l|cc|c|c|c|c}
\hline
\textbf{Article} & \multicolumn{2}{c|}{Liu  \cite{Liu}} & Janveja \cite{Janveja} & Hu \cite{Hu} & \textit{Proposed} \\ \hline
\textbf{Dataset} & \multicolumn{2}{c|}{MIT-BIH} & MIT-BIH & MIT-BIH & \textit{MIT-BIH} \\ \hline
\textbf{Model Type} & \multicolumn{1}{c|}{DNN} & CNN & DNN & DBLSTM & \textit{DNN} \\ 
\textbf{Accuracy (\%)} & \multicolumn{1}{c|}{99.30} & 99.16 & 91.6 & 96.74 & \textit{93.68} \\ 
\textbf{Sensitivity (\%)} & \multicolumn{1}{c|}{99.30} & \begin{tabular}[c]{@{}l@{}}98.2 (V)\\ 87.30 (S)\end{tabular} & \begin{tabular}[c]{@{}l@{}}83.37 (V) \\ 86.45 (S)\end{tabular} & - & \textit{92.08} \\ 
\begin{tabular}[c]{@{}l@{}}\textbf{Model Size} \\ (\textbf{Trainable} \\ \textbf{Parameters})\end{tabular} & \multicolumn{1}{c|}{90352} & 14117 & 8415 & 8604 & \textit{8415} \\ 
\textbf{Weight precision} & \multicolumn{1}{c|}{16INT} & 16INT & 16INT & 4INT  & \textit{8INT} \\ 
\textbf{Data precision} & \multicolumn{1}{c|}{16INT} & 16INT & 16INT & 8INT & 8INT \\ 
\textbf{Classes} & \multicolumn{1}{c|}{2} & 5 & 5 & 5 & \textit{5} \\ 
\textbf{Implementation} & \multicolumn{1}{c|}{ASIC} & ASIC & ASIC & FPGA & \textit{ASIC} \\ 
\textbf{Power ($\mu$W)} & \multicolumn{1}{c|}{46.8} & 86.7 & 8.75 & 40 (Dynamic) & \textit{3.07} \\ 
\textbf{Area (mm$^{2}$)} & \multicolumn{1}{c|}{1.74} & 1.74 & 1.32 & - & 1.03 \\ 
\textbf{Frequency} & \multicolumn{1}{c|}{1 Mhz} & 2.5 MHz & 12 kHz & 40 kHz & \textit{12 kHz} \\ 
\textbf{Process} & \multicolumn{1}{c|}{65nm} & 65nm & 180nm SCL & Artix-7 & \textit{180nm UMC} \\ 
\textbf{Energy/Class ($\mu$J)} & \multicolumn{1}{c|}{2.25} & 4.36 & 2.08 & - & \textit{2.17} \\ \hline
\end{tabular}

\end{table*}

\section{Discussion}


\textit{Peak Detection:}
Accuracy and sensitivity using inputs segmented with the peak detection are higher on average  than when the annotated reference peaks are used, even though some peaks are missed or inaccurately located by the algorithm. The S class is particularly affected, with sensitivity dropping from over 92\% down to 85\%. This can be explained by the fact that the model is trained on data segmented using the peak detection algorithm. This indicates that the model is highly sensitive to peak detection shifts. Consequently, the peak detection algorithm may introduce systematic offsets in peak locations; however, this does not affect classification accuracy, since the model is trained on data with the same offsets, resulting in strong performance regardless. When the annotated reference peaks are used instead, they introduce a distributional mismatch with the training data, which explains the lower accuracy.

\textit{16-bit vs 8-bit Architectures:}
As shown in Table \ref{tab:litcomp}, our proposed 8-bit model provides a better classification accuracy and sensitivity than the 16-bit reference model. This aligns with findings in the literature, where reduced precision has been observed to lower the risk of overfitting by acting as a form of implicit regularization \cite{Ansari}, an an effect that is confirmed by the results of this work.

\textit{Error Profiles of Approximate Multipliers:}
Approximate multipliers with more symmetrical error distributions, such as CSSM4 and ILM-0, achieved higher classification accuracy in the neural network compared to multipliers with one-sided error distributions, such as SSM4. Symmetrical errors are more likely to cancel out across successive multiply-accumulate operations within the network, whereas one-sided errors accumulate and introduce a systematic bias that shifts activations away from their expected values, ultimately degrading classification performance.

\textit{The ILM-0 Multiplier:}
The ILM-0 multiplier did perform well when tested for classification accuracy, however when it came to power usage it was outperformed by the accurate multiplier. The original paper that proposed the ILM-0 multiplier evaluated power at a clock frequency of 250 MHz, whereas in this work it operates at 12 kHz. At such a low frequency, the timing constraints are easily met regardless of multiplier architecture, negating the power advantage that ILM-0 offers at higher frequencies.

\textit{Implications for Clinical Practice:}
The results demonstrate that approximate computing techniques substantially reduce power consumption in DNN-based arrhythmia classification with only a marginal decrease in accuracy and sensitivity. This is particularly relevant for wearable ECG monitoring devices, where lower power consumption enables longer battery life and continuous long-term monitoring. The strong performance of the 8-bit model further suggests that lightweight, low-precision models are viable for real-time, on-device arrhythmia detection. Importantly, the classification accuracy achieved by the proposed approximate designs exceeds the requirements reported in \cite{hannun2019cardiologist}, confirming their suitability for deployment in wearable ECG monitoring devices. Such devices serve as screening tools that provide early indications of potential cardiac abnormalities, prompting the wearer to seek professional medical evaluation. In this context, the power savings enabled by approximate multipliers are particularly valuable, as they extend battery life and support the continuous long-term monitoring that is essential for detecting intermittent arrhythmias. To further strengthen clinical applicability, future work could explore class-specific optimization of approximate multiplier selection, ensuring that power savings are maximized while maintaining adequate sensitivity for each arrhythmia class.


\section{Conclusion}

In this work, we have investigated approximate computing techniques for DL based low-power and energy-efficient arrhythmia detection suitable for wearable devices. The goal is to achieve a higher power/energy efficiency compared to that of the state-of-the-art architectures while maintaining a classification sensitivity higher than that of an average cardiologist, i.e., 78.4\% \cite{hannun2019cardiologist}. Specifically, 
data precision reduction and approximate multiplication have been investigated, which includes approximation-aware retraining. Our proposed 8-bit architecture (S8GC(R)) has shown an improvement of 64.9\% in power efficiency as compared to the state-of-the-art architecture while providing an acceptable classification performance (93.7\% classification accuracy
and 92.1\% sensitivity). Moreover, we have investigated a relatively higher frequency of operation (100 MHz) to improve energy efficiency, which provided an improvement of 61.5\% compared to the state-of-the-art architecture. Therefore, our proposed architecture is a significant step towards improving the battery life of wearable devices that can detect arrhythmia within the required classification performance.

\bibliographystyle{IEEEtran}
\bibliography{bibliography}
\end{document}